\documentclass{epl2}

\usepackage{epsfig,ams,amsmath}

\title{Charge and spin Hall effect in graphene with magnetic impurities}

\author{K.-H. Ding,\inst{1} Z.-G. Zhu,\inst{2}  J. Berakdar\inst{2}}
\shortauthor{Ding \etal}

\institute{ \inst{1} Department of Physics and Electronic Science,
 Changsha University of Science and Technology,
 Changsha 410076, China\\
  \inst{2} Institut f\"ur Physik, Martin-Luther-Universit\"at
 Halle-Wittenberg,   Heinrich-Damerow-St. 4, 06120 Halle, Germany }
\abstract{We point out the existence of   finite  charge  and spin Hall conductivities  of
graphene in the presence of a spin orbit interaction (SOI)
and localized  magnetic impurities.
The SOI in graphene results in different transverse forces on
the two  spin channels yielding the spin Hall current.  The
magnetic scatterers act as spin-dependent barriers, and in
combination  with the SOI effect lead to a charge imbalance at the boundaries.
As indicated here, the charge and spin Hall effects should be  observable in
graphene by changing the chemical potential close to the gap.}

\begin{document}
 \maketitle

\section{Introduction} Graphene,
 currently under intense experimental and
theoretical investigations \cite{novoselov2,zhang,geim,castrocond},
  consists of a single layer of  carbon atoms arranged in a honeycomb lattice made of
  two interpenetrating, $\sigma$-bonded triangular sublattices, $A$ and $B$.
The low-energy spectrum and the transport properties near the neutrality point are dictated by
  the $\pi$ and $\pi^*$  bands that form
  conical valleys touching  at the high symmetry points $K$ and $K'$ of the
   Brillouin zone \cite{McClureprb1956}.
  A key point is the linear dispersion near  $K$ and $K'$
   which renders a low-energy theory in terms of an
 effective Dirac-Weyl Hamiltonian \cite{castrocond}.  
This results in a number of fascinating   phenomena such as
the half-quantized Hall effect
\cite{novoselov2,zhang,gusyninprl2005,netoprb2006}. Remarkably,
  the spin relaxation length is measured to be as long as
$\sim$ 1.5 $\mu m$ in low-mobility devices \cite{4} which makes
graphene interesting for spintronic applications. In this regard
the  spin-orbit coupling (SOI) \cite{5} is a decisive  factor as
it determines the spin-decoherence time and opens the way  for the
spin manipulation and control. Generally, in ferromagnets  SOI
plays a key role in the anomalous Hall effect (AHE)
\cite{6,mitphys1955,8,9,nagaosa0904,sinitsynjpc2008} and it is
also essential for the spin Hall effect (SHE)
\cite{dyakonovpl1971,hirschprl1999,murakamisci2003,sinovaprl2004}.
 Manifestations of  SOI in graphene have been addressed by
Kane \emph{et al.}\cite{kaneprl20051,kaneprl20052} indicating that
 the spin Hall conductivity in the undoped graphene is
quantized due to a gap produced by SOIs in the absence of a
magnetic field. Following Kane \emph{et al.}'s work, Sinitsyn \emph{et
al.} \cite{sinitsynprl2006} predicted  a substantial spin Hall
conductivity in doped graphene due to the skew
scattering, and find that the charge Hall conductivity vanishes
because of a cancelation of contributions of bands   with  opposite spins.

In this work, we study the  Hall effect in graphene in
the presence of SOI and localized magnetic impurities. We find the
charge Hall effect is generally finite as a result of the
combined influence of  SOI and spin flip scattering at the magnetic
impurities, a mechanism different from   the skew
scattering\cite{mitphys1955} and the side jump \cite{8}
mechanisms established for AHE in  ferromagnetic metals. The SOI in graphene  exerts
an asymmetric transverse force on electrons with  different  spins,
generating thus a spin Hall current. In addition, the scattering off the
magnetic impurities serves  as  a spin dependent barrier leading to
a charge imbalance at the boundary of the sample.
We  argue that the charge and the spin Hall
effects should be observable  in the presence of
SOI and magnetic impurities by changing the chemical potential
close to the gap.

\section{Theoretical model} The Hamiltonian for the clean
undoped graphene with SOI  is  \cite{kaneprl20051,ding08}
\begin{equation}
H=\sum\limits_{\mathbf{k}}\psi_\mathbf{k}^\dag(
v_F\mathbf{k}\cdot\mathbf{\sigma}\tau_z+\Delta\sigma_z s_z)\psi_\mathbf{k},
\label{h}
\end{equation}
where $\psi_\mathbf{k}^\dag\equiv(c_{\mathbf{k}AK\uparrow}^\dag,$
$c_{\mathbf{k}BK\uparrow}^\dag, c_{\mathbf{k}BK'\uparrow}^\dag,
c_{\mathbf{k}AK'\uparrow}^\dag, c_{\mathbf{k}AK\downarrow}^\dag,$
$c_{\mathbf{k}BK\downarrow}^\dag,c_{\mathbf{k}BK'\downarrow}^\dag,c_{\mathbf{k}AK'\downarrow}^\dag)$.
  $c_{\mathbf{k}\sigma\tau s}^\dag$ is the  creation operator of a single-particle state
   labeled by the   momentum $\mathbf{k}$ and  associated with the sublattice $\sigma\equiv A,B$ and the Dirac point
 $\tau\equiv K,K'$, and having the
 spin  $s=\uparrow,\downarrow$.  $v_F$ is the Fermi velocity  and
 $\Delta$ is the SOI strength parameter. Upon
 doping with localized magnetic impurities the  graphene carriers
 couple to the impurities via
\begin{equation}
H_i=\frac{1}{V}\sum\limits_{\mathbf{k},\mathbf{q}}\psi_{\mathbf{k}+\mathbf{q}}^\dag
V_i(\mathbf{q})\psi_\mathbf{k}, \label{h_i}
\end{equation}
where $V$ is the volume of the system, and
\begin{equation}
V_i(\mathbf{q})=\left(
\begin{array}{cc}
u_1\rho_i(\mathbf{q})&u_2\sigma_x\rho_i(\mathbf{Q}+\mathbf{q})\\
u_2\sigma_x\rho_i(-\mathbf{Q}+\mathbf{q})&u_1\rho_i(\mathbf{q})
\end{array}\right)\mathbf{s}\cdot
\mathbf{S}.
 \label{v_i}\end{equation}
 $\rho_i(\mathbf{q})$ stand for the Fourier components of the
impurity density, $\mathbf{Q}$ is a vector pointing from the point
$K$ to the point $K'$, $u_1$ and $u_2$ are the intra-valley and
inter-valley scattering strengths, respectively, and $\mathbf{S}$
is the spin operator of the localized magnetic impurities
\cite{ave}. The case of a single localized impurity has been
treated in \cite{uchi}.
To obtain the spin-dependent lifetimes of the quasi-particles and
their energetic positions in the presence of  impurities we
calculate the spin-dependent imaginary and real parts of the
self-energies $\Sigma$, which in the Born approximation \cite{kon}
reads
\begin{equation}
\begin{array}{cll}
\Sigma(\mathbf{k},\mathbf{k}';\omega+i\eta)
=\frac{1}{V^2}\sum\limits_{\mathbf{q}} \langle
V_i(\mathbf{k}-\mathbf{q})\mathcal{G}_0({\mathbf{q}};\omega+i\eta)
V_i(\mathbf{q}-\mathbf{k}')\rangle
\end{array} \end{equation}
where $\langle ...\rangle$ stands for the impurity average and $\mathcal{G}_0$ is the resolvent of $H$.
Straightforward calculations show that in spin space $\Sigma$ has the structure
\begin{equation}
\begin{array}{cll}
\Sigma(\omega+i\eta)=
\left(
\begin{array}{cc}
\Sigma_0^\uparrow+i\Sigma_c^\uparrow\sigma_z&0\\
0&\Sigma_0^\downarrow+i\Sigma_c^\downarrow\sigma_z
\end{array}\right).
\end{array}
\label{selfenergy}\end{equation}
 The matrix
expressions for the impurity averaged, retarded (R) or advanced
(A) Green functions  are \cite{kon1}
\begin{equation}
\begin{array}{cll}
G^{R/A}(\mathbf{k},\omega) &=&\left(
\begin{array}{cc}
\frac{\omega-\Sigma_0^{\uparrow R/A}+v_F\mathbf{k}\cdot
\sigma\tau_z+(\Delta+\Sigma_c^{\uparrow R/A})\sigma_z}{(\omega-
\Sigma_0^{\uparrow R/A})^2-v_F^2\mathbf{k}^2-(\Delta+\Sigma_c^{\uparrow R/A})^2}&0\\
0&\frac{\omega-\Sigma_0^{\downarrow R/A}+v_F\mathbf{k}\cdot
\sigma\tau_z-(\Delta-\Sigma_c^{\downarrow R/A})\sigma_z}
{(\omega-\Sigma_0^{\downarrow R/A})^2-v_F^2\mathbf{k}^2-(\Delta-\Sigma_c^{\downarrow R/A})^2}
\end{array}\right).
\end{array}\label{grn}
\end{equation}
From (\ref{selfenergy}) it is obvious that $\Sigma$ consists of a
spin-dependent but sub-lattice independent part $\Sigma_0$,
whereas $\Sigma_c$ depends on both the sublattice index and on the
spin. Thereby we find the explicit expressions
$$\Sigma_0^{s R/A}=\Lambda_0^s\pm
i\Gamma_0^s,\ \
\Sigma_c^{s R/A}=\Lambda_c^s\pm
i\Gamma_c^s,
$$
where
$$
\Lambda_0^s=-\frac{n_i(u_1^2+u_2^2)(
a_s  +\langle S_z^2\rangle)\omega}{4\pi
v_F^2}\ln|\frac{\omega^2-\Delta^2-D^2} {\omega^2-\Delta^2}|,$$ and
$$\Gamma_0^s=-\frac{\theta(0<\omega^2-\Delta^2<D^2)\pi
n_i(u_1^2+u_2^2)(a_s+\langle
S_z^2\rangle)|\omega|}{4\pi v_F^2},$$
$$ \Lambda_c^s=\frac{(-1)^sn_i(u_1^2-u_2^2)(a_s-\langle
S_z^2\rangle)\Delta}{4\pi v_F^2}\ln|\frac{\omega^2-\Delta^2-D^2}
{\omega^2-\Delta^2}|,$$ and
$$\Gamma_c^s=\frac{(-1)^s\theta(0<\omega^2-\Delta^2<D^2)\text{sgn}(\omega)\pi
n_i(u_1^2-u_2^2)(a_s-\langle S_z^2\rangle)\Delta}{4\pi v_F^2}.$$
Here, $$a_{s=\uparrow}=\langle S_-S_+\rangle,\, a_{s=\downarrow}=\langle S_+S_-\rangle,$$
 $(-1)^s=\pm 1$ for $s=\uparrow \downarrow$, and  $S_\pm =S_x\pm iS_y$
 are the lowering and raising operators of the impurity spin.
 $\theta(x)$ is the step function, and  $n_i$ is the average
impurity density. $D=v_Fk_c$ and  $k_c$  is a cutoff
for the $k$ summation.

\section{Charge and spin currents}
The $i$ component of the charge current operator is
 $$ J_i=e\frac{\partial H}{\partial
k_i}=ev_F\sigma_i\tau_z.$$ The spin current with the spin being
along the $z$ axis and flowing in the x direction  we evaluate
from the anti-commutator between the velocity operator and the
Pauli matrix\cite{sinovaprl2004}
\begin{equation}
J_x^z=\frac{\hbar}{4}\{s_z,\frac{\partial H}{\partial
k_x}\}=\frac{\hbar}{2}v_F\sigma_x\tau_zs_z.
\end{equation}
 The electric conductivity $\sigma_{ij}$ is the (Kubo) linear
response function to the external electric field \cite{mahan1990}
 $$
\sigma_{ij}=\lim\limits_{\Omega\rightarrow
0}\frac{\text{Im}\Pi_{ij}^R(\Omega+i\eta)}{\Omega},
$$
where $\Pi_{ij}^R(\omega)$ is the retarded current-current
correlation function obtained from an analytic continuation of the
Matsubara function

 $$
\Pi_{ij}(i\Omega_n)=\frac{1}{V}\int_0^\beta d\tau
e^{i\Omega_n\tau}\langle {\cal T}_\tau J_i(\tau)J_j(0)\rangle,\ \
\Omega_n=2\pi nT
$$
and $\beta$ is the inverse temperature ($T$) and ${\cal T}_\tau$ is the $\tau$ ordering operator.
 We calculated $\sigma_{ij}$
by adopting the ladder approximation for the current vertex
determined by Bethe-Salpeter equation \cite{inoue04,inoue06}.
\begin{equation}
\Gamma_x^{\lambda\lambda'}(\omega,\omega')=\sigma_x\tau_z+\frac{1}{V^2}\sum\limits_{\mathbf{k}'}
\langle V_{i}(\mathbf{k}-\mathbf{k}')G^\lambda(\mathbf{k}',\omega)
\Gamma_x^{\lambda\lambda'}(\omega,\omega')G^{\lambda'}(\mathbf{k}',\omega')V_{i}(\mathbf{k}'-\mathbf{k})\rangle,\label{gammax}
\end{equation}
where  $\lambda,\lambda'=R,A$; $\omega'=\omega+\Omega$ and
$\langle\cdots\rangle$ stand for the average over the impurity
distributions.
By iteration we express  Eq.(\ref{gammax})
as ($\Gamma_x^{\lambda\lambda'}\equiv\Gamma_x^{\lambda\lambda'}(\omega,\omega')$)
\begin{equation}
\begin{array}{cll}
\Gamma_x^{\lambda\lambda'}
&=&a^{\lambda\lambda'}\sigma_x\tau_z+b^{\lambda\lambda'}\sigma_y\tau_z+c^{\lambda\lambda'}\sigma_x\tau_zs_z
+d^{\lambda\lambda'}\sigma_y\tau_zs_z .\label{gaiter}
\end{array}
\end{equation}
 Substituting Eq.(\ref{gaiter}) in Eq.(\ref{gammax}) we find
\begin{eqnarray}
a^{\lambda\lambda'}+c^{\lambda\lambda'}
 &=& 1+\alpha_0\langle
S_-S_+\rangle \Pi_2[\Xi_3(a^{\lambda\lambda'}-c^{\lambda\lambda'})
+i\Xi_4(b^{\lambda\lambda'}-d^{\lambda\lambda'})]
+\alpha_0\langle S_z^2\rangle \nonumber\\
&& \Pi_1[\Xi_1(a^{\lambda\lambda'}+c^{\lambda\lambda'})-i\Xi_2(b^{\lambda\lambda'}+d^{\lambda\lambda'})],
\label{eg1}
\end{eqnarray}
%
%
%
\begin{equation}
\begin{array}{cll}
a^{\lambda\lambda'}-c^{\lambda\lambda'}
&=&1+\alpha_0\langle S_z^2\rangle
\Pi_2[\Xi_3(a^{\lambda\lambda'}-c^{\lambda\lambda'})+i\Xi_4(b^{\lambda\lambda'}-d^{\lambda\lambda'})]
+\alpha_0\langle S_+S_-\rangle\nonumber\\
&&\Pi_1[\Xi_1(a^{\lambda\lambda'}+c^{\lambda\lambda'})-i\Xi_2(b^{\lambda\lambda'}+d^{\lambda\lambda'})],
\end{array}\label{eg2}
\end{equation}
%
%
%
\begin{equation}
\begin{array}{cll}
b^{\lambda\lambda'}+d^{\lambda\lambda'}
&=&\beta_0\langle S_-S_+\rangle
\Pi_2[\Xi_3(b^{\lambda\lambda'}-d^{\lambda\lambda'})-i\Xi_4(a^{\lambda\lambda'}-c^{\lambda\lambda'})]
+\beta_0\langle S_z^2\rangle\nonumber\\
&&\Pi_1[\Xi_1(b^{\lambda\lambda'}+d^{\lambda\lambda'})+i\Xi_2(a^{\lambda\lambda'}+c^{\lambda\lambda'})],
\end{array}\label{eg3}
\end{equation}
\begin{equation}
\begin{array}{cll}
b^{\lambda\lambda'}-d^{\lambda\lambda'}
&=&\beta_0\langle S_z^2\rangle
\Pi_2[\Xi_3(b^{\lambda\lambda'}-d^{\lambda\lambda'})-i\Xi_4(a^{\lambda\lambda'}-c^{\lambda\lambda'})]
+\beta_0\langle S_+S_-\rangle\nonumber\\
&&\Pi_1[\Xi_1(b^{\lambda\lambda'}+d^{\lambda\lambda'})+i\Xi_2(a^{\lambda\lambda'}+c^{\lambda\lambda'})],
\end{array}\label{eg4}
\end{equation}
%
%
%
%
where $\alpha_0=\frac{n_i(u_1^2-u_2^2)}{4\pi v_F^2}$,
$\beta_0=\frac{n_i(u_1^2+u_2^2)}{4\pi v_F^2}$,
$\Pi_i\equiv\Pi_i^{\lambda\lambda'}$,
$\Xi_i\equiv\Xi_i^{\lambda\lambda'}$, and
%
$$
\begin{array}{cll}
\Pi_{1(2)}^{\lambda\lambda'}=\frac{\ln\frac{(\omega-\Sigma_0^{\uparrow(\downarrow),\lambda})^2
-(\Delta\pm\Sigma_c^{\uparrow(\downarrow),\lambda})^2}{(\omega'-\Sigma_0^{\uparrow(\downarrow),\lambda'})^2
-(\Delta\pm\Sigma_c^{\uparrow(\downarrow),\lambda'})^2}}{(\omega'-\Sigma_0^{\uparrow(\downarrow),\lambda'})^2
-(\omega-\Sigma_0^{\uparrow(\downarrow),\lambda})^2
-(\Delta\pm\Sigma_c^{\uparrow(\downarrow),\lambda'})^2+(\Delta\pm\Sigma_c^{\uparrow(\downarrow),\lambda})^2},
\end{array}
$$
%
$$
\Xi_{1(3)}^{\lambda\lambda'}=(\omega-\Sigma_0^{\uparrow(\downarrow),\lambda})(\omega'-\Sigma_0^{\uparrow(\downarrow),\lambda'})
-(\Delta\pm\Sigma_c^{\uparrow(\downarrow),\lambda})(\Delta\pm\Sigma_c^{\uparrow(\downarrow),\lambda'}),$$
%
%
$$
\Xi_{2(4)}^{\lambda\lambda'}=(\Delta\pm\Sigma_c^{\uparrow(\downarrow),\lambda})(\omega'-\Sigma_0^{\uparrow(\downarrow),\lambda'})
-(\omega-\Sigma_0^{\uparrow(\downarrow),\lambda})(\Delta\pm\Sigma_c^{\uparrow(\downarrow),\lambda'}),
$$
Using  Eq.(\ref{gaiter}) we obtain for
the Hall conductivity at zero temperature the first central result
\begin{equation}
\begin{array}{cll}
\sigma_{yx} = \frac{e^2}{ 2\pi^2} \left\{
T_1(b^{AR}+d^{AR})+T_2(b^{AR}-d^{AR})
+T_3(a^{AR}+c^{AR})+T_4(a^{AR}-c^{AR}) \right\} ,\label{syx}
\end{array}
\end{equation}
where $a^{AR},b^{AR},c^{AR},d^{AR}$ are the solutions of
Eqs.(\ref{eg1})-(\ref{eg4}) in the limit $\Omega\rightarrow0$, and ($\mu$ is the chemical potential)
\begin{equation}
\begin{array}{cll}
T_{1(2)}
=\frac{1}{2}\frac{(\mu-\Lambda_0^{\uparrow(\downarrow)})^2-(\Delta\pm\Lambda_c^{\uparrow(\downarrow)})^2+
\Gamma_0^{\uparrow(\downarrow)2}
-\Gamma_c^{\uparrow(\downarrow)2}}{(\mu-\Lambda_0^{\uparrow(\downarrow)})\Gamma_0^{\uparrow(\downarrow)}
\pm(\Delta\pm\Lambda_c^{\uparrow(\downarrow)})\Gamma_c^{\uparrow(\downarrow)}}
\:\arctan\frac{(\mu-\Lambda_0^{\uparrow(\downarrow)})^2-\Gamma_0^{\uparrow(\downarrow)2}-
(\Delta\pm\Lambda_c^{\uparrow(\downarrow)})^2
+\Gamma_c^{\uparrow(\downarrow)2}}
{2(\mu-\Lambda_0^{\uparrow(\downarrow)})\Gamma_0^{\uparrow(\downarrow)}\pm2(\Delta\pm\Lambda_c^{\uparrow(\downarrow)})
\Gamma_c^{\uparrow(\downarrow)}},\\
\end{array}
\end{equation}
\begin{equation}
\begin{array}{cll}
T_{{3}{(4)}} =\frac{\mu\Gamma_c^{\uparrow(\downarrow)}\pm\Delta\Gamma_0^{\uparrow(\downarrow)}
+(\Lambda_c^{\uparrow(\downarrow)}\Gamma_0^{\uparrow(\downarrow)}-\Lambda_0^{\uparrow(\downarrow)}
\Gamma_c^{\uparrow(\downarrow)})}
{(\mu-\Lambda_0^{\uparrow(\downarrow)})\Gamma_0^{\uparrow(\downarrow)}
\pm(\Delta\pm\Lambda_c^{\uparrow(\downarrow)})\Gamma_c^{\uparrow(\downarrow)}}
\:\arctan\frac{(\mu-\Lambda_0^{\uparrow(\downarrow)})^2-
\Gamma_0^{\uparrow(\downarrow)2}-(\Delta\pm\Lambda_c^{\uparrow(\downarrow)})^2
+\Gamma_c^{\uparrow(\downarrow)2}}
{2(\mu-\Lambda_0^{\uparrow(\downarrow)})\Gamma_0^{\uparrow(\downarrow)}
\pm2(\Delta\pm\Lambda_c^{\uparrow(\downarrow)})\Gamma_c^{\uparrow(\downarrow)}}.
\end{array}\label{t34}
\end{equation}
From this procedure we obtain for the spin conductivity
\begin{equation}
\begin{array}{cll}
\sigma_{xx}^s = \frac{e\hbar}{ 4\pi^2} \{ T_1(a^{AR}+c^{AR}) -
T_2(a^{AR}-c^{AR})
-T_3(b^{AR}+d^{AR})+T_4(b^{AR}-d^{AR})+C \},
\end{array}
\end{equation}
where
$C =\frac{\alpha_0(\langle S_+S_-\rangle-\langle
S_-S_+\rangle)}{(1+\alpha_0\langle
S_z^2\rangle)^2-\alpha_0^2\langle S_-S_+\rangle\langle
S_+S_-\rangle}.$
 The spin Hall conductivity is
\begin{equation}
\begin{array}{cll}
\sigma_{yx}^s =\frac{e\hbar}{ 4\pi^2} \{T_1(b^{AR}+d^{AR})
-T_2(b^{AR}-d^{AR})
+T_3(a^{AR}+c^{AR}) -T_4(a^{AR}-c^{AR})\}.
\end{array}
\label{spinhall}
\end{equation}

\section{Discussions}
From the above derivation the following
 general conclusions are inferred:
  \textbf{\emph{i})} When the SOI strength $\Delta=0$ and
  the impurity spin ladder operators
$S^+,S^-\neq0$, we find for the sublattice-dependent part of the
self-energy
 $\Sigma_c^{\uparrow R/A}=\Sigma_c^{\downarrow R/A}=0$,
the parameters in Eq.(\ref{t34}) $T_3=T_4=0$, and the parameters
in Eq.(\ref{eg1}) $\Xi_2^{AR}=\Xi_4^{AR}=0$. Substituting these
results in Eqs.(\ref{eg1})-(\ref{eg4}), we obtain
$b^{AR}-d^{AR}=b^{AR}+d^{AR}=0$. This means that the charge Hall
conductivity $\sigma_{yx}=0$  vanishes (see Eq.(\ref{syx})) as
well as the spin Hall conductivity $\sigma_{yx}^s=0$ (see
Eq.(\ref{spinhall})). Hence, we conclude that merely spin-flips at
the magnetic impurities do not  induce charge and/or spin Hall
effect in absence of  SOI. \textbf{\emph{ii})} For \emph{the SOI
strength} $\Delta\neq0$ and the impurity spin operators
 $S^+=S^-=0$, our model for the magnetic impurities is not
reducible to the nonmagnetic impurity case;  the former case is
related to the Pauli spin matrix $s_{z}$, while the latter is
expressed in a unit matrix, but they show  similar behaviour.
Because in our case
 $a^{AR}-c^{AR}=a^{AR}+c^{AR}$ and
$b^{AR}-d^{AR}=-(b^{AR}+d^{AR})$ when the self-energies satisfy
the relations: $\Sigma_0^{\uparrow R/A}=\Sigma_0^{\downarrow R/A}$
and $\Sigma_c^{\uparrow R/A}=-\Sigma_c^{\downarrow R/A}$, we
conclude that the charge Hall conductivity $\sigma_{yx}=0$ and the
spin Hall conductivity is not vanishing; a result  which is
consistent with the findings of Ref.\cite{sinitsynprl2006}. The
charge Hall conductivity vanishes because of a cancelation between
bands with opposite spins. \textbf{\emph{iii})} For the SOI
strength $\Delta\neq0$ and the impurity spin operators
$S^+,S^-\neq0$, the charge and spin Hall conductivities are in
general nonvanishing, similarly  as in the case of magnetic
impurities
 in two dimensional electron gas with Rashba SOI \cite{inoue06,wang07}. However,
in the latter case, the spin Hall conductivity is suppressed to
zero by scattering from nonmagnetic impurities with a linear
 Rashba SOI \cite{rashbaimp}.

This analysis suggests that the charge Hall effect is the
consequence of the combined influence of the SOI and the spin flip
scattering at the magnetic impurities. This  is an essentially
different mechanism from the skew scattering \cite{mitphys1955}
and the side jump mechanism \cite{8} in the ferromagnetic metals.
In fact, the latter do not occur in our graphene model since the
SOI is intrinsic and homogenous\cite{soi}. Evidence is provided by
investigating the longitudinal spin conductivity. For example,
when $\Delta=0$ and $S^+,S^-\neq0$ it is straightforward to verify
that the spin Hall conductivity vanishes, but
$\sigma_{xx}^s\neq0$, meaning that the scattering of spin-up and
spin-down electrons by magnetic impurities in graphene is
different  in the longitudinal direction only, but does not yield
an effective force in the transverse direction, thereby resulting
in a non-vanishing spin current and $\sigma_{yx}^s=0$. In
contrast, for $\Delta\neq0$ and $S^+,S^-=0$, we obtain
$\sigma_{yx}^s\neq 0$ and $\sigma_{xx}^s=0$. This is due to
opposite SOI forces in transverse direction on the charge current
of the spin-up and spin-down electrons. Thus, we conclude that the
presence of both the SOI and the magnetic impurity  brings about
the occurrence of the charge Hall effect without external magnetic
field in graphene.

In the Boltzmann limit\cite{zhengprb2002}, which in our model  is
achieved when $\Lambda_0^s,\Lambda_c^s=0$ and
$|\mu|\gg|\Gamma_{0/c}^\sigma|$  we find
\begin{equation}
\sigma_{yx}=-\frac{4e^2}{\pi}\frac{(\langle S_+S_-\rangle-\langle
S_z^2\rangle)(\langle S_-S_+\rangle-\langle S_z^2\rangle)(\langle
S_+S_-\rangle^2-\langle S_-S_+\rangle^2) }{(\langle
S_z^2\rangle^2-\langle S_-S_+\rangle\langle S_+S_-\rangle)^2
}\frac{\Delta}{ |\mu|},\label{sigmayx1}
\end{equation}
\begin{equation}
\sigma_{yx}^s =-\frac{2e\hbar}{\pi}\frac{(\langle
S_+S_-\rangle+\langle S_z^2\rangle)(\langle S_-S_+\rangle+\langle
S_z^2\rangle)[ (\langle S_+S_-\rangle-\langle
S_z^2\rangle)^2+(\langle S_-S_+\rangle-\langle S_z^2\rangle)^2]
}{(\langle S_z^2\rangle^2-\langle S_-S_+\rangle\langle
S_+S_-\rangle)^2 }\frac{\Delta}{ |\mu|}.\label{sigmayx2}
\end{equation}
Both the charge Hall conductivity and the spin Hall conductivity
are inversely proportional to the absolute value of the chemical
potential, and independent of the electron-impurity interaction
strength and the concentration of scatterers. From
Eqs.(\ref{sigmayx1}) and (\ref{sigmayx2}) we deduce that the
charge and spin Hall effect should in principle be observable in
graphene doped  with magnetic impurities by changing the chemical
potential close to the gap. As for the presence of SOI in graphene, we note that in recent
experiments  graphene is deposited on top of a
Ni(111) substrate. It is argued that the formation of a charge density gradient in
the interface layer results in the experimentally observed large Rashba splitting
in graphene \cite{yuprl2008}. Another aspect is that  intercalated Au
provides a $\approx$100-fold enhancement of the spin-orbit
splitting of graphene $\pi$ states \cite{Varykhalovprl2008}. In addition, as an open surface, graphene
 allows precise  adatoms manipulations, as recently demonstrated experimentally \cite{meyer}.
Furthermore, when graphene is deposited on a ferromagnetic
material such as Ni, the coupling between the Dirac fermion in
graphene and the magnetic atom in the substrate  arises
inevitably due to roughness in the underlying substrate surface
\cite{yuanboarxiv,weserarxiv}. Motivated by these facts, we are
hopeful that an possible experimental realization of
charge and spin Hall effect in graphene in the presence of
magnetic impurities, as uncovered here, should be possible.

\section{Summary} We studied the charge and spin Hall effects in
graphene with magnetic impurities. Using the Kubo formula, we
obtained  analytical expressions for the charge and spin Hall conductivities
and concluded that both are generally finite
in the presence of the SOI and magnetic impurities under zero
external magnetic field. The charge Hall effect originates from
a combined action of SOI and spin flip scattering at the
magnetic impurities.
The SOI in graphene results in  transverse forces different for
the two  spin channels which yields a  spin Hall current. On the other
hand, the scattering from the magnetic impurities act as a
spin-dependent barrier causing the imbalance of the charge
accumulation at the boundaries. The derived results for the charge and
spin Hall effects should be observable by current
technology \cite{meyer}
 by changing the chemical potential close to
the gap.

The work of K.H.D. was supported by the National Natural Science
Foundation of China (Grant No. 10904007), the Natural Science
Foundation of Hunan Province, China (Grant No. 08JJ4002 ), and the
construct program of the key discipline in Changsha University of
Science and Technology, China. J.B. and Z.G.Z. are supported by
the cluster of excellence "Nanostructured Materials" of the state
Saxony-Anhalt, Germany.


\begin{thebibliography}{19}
\bibitem{kon} We operate in the regime where the Kondo effect is subsidiary.
\bibitem{kon1} Note, this representation of the Green's function is in the
spin space.
\bibitem{soi}   SOI has various sources in graphene, e.g., due to the  specific honeycomb
lattice structure  only the normal component has a
nonzero contribution due to the reflection symmetry with respect
to the lattice plane and is not dependent on the momentum of the
electrons in contrast to a convenetional two-dimensional electron gas.
\end{thebibliography}
\end{document}